# Gigahertz Dielectric Polarization of Single-atom Niobium Substituted in Graphitic layers

*Xuefeng Zhang,* [1, 2*] *Junjie Guo,* [3] *Pengfei Guan,* [4] *Gaowu Qin,* [1*] *Stephen J. Pennycook* [3,5]

[1] *Key Laboratory for Anisotropy and Texture of Materials (MOE), Northeastern University, Shenyang 110819, P. R. China*

[2] *National Research Council Canada, 75, de Mortagne, Boucherville, Québec, J4B 6Y4, Canada*

[3] *Materials Science and Technology Division, Oak Ridge National Laboratory, Oak Ridge, Tennessee 37831, USA*

[4] *Beijing Computational Science Research Center, Beijing 100084, China*

[5] *Dept. of Materials Science and Engineering, National University of Singapore, Singapore 117576*

**We have synthesized two Nb@C composites with an order of magnitude difference in the density of single-atom niobium substituted into graphitic layers. The concentration and sites of single-atom Nb are identified using aberration-corrected scanning transmission electron microscopy and density functional theory. Comparing the complex permittivity spectra show that the representative dielectric resonance at ~16 GHz originates from the intrinsic polarization of single-atom Nb sites, confirmed by theoretical simulations. The single-atom dielectric resonance represents the physical limit of the electromagnetic response of condensed matter, and thus might open up a new avenue for designing electromagnetic wave absorption materials. Single-atom resonance also has important**

**implications in understanding the correlation between the macroscopic dielectric behaviors and the atomic-scale structural origin.**

In solids, dielectric phenomena are induced by vibration processes of asymmetric electric charges or dipoles, arising from the polarization of structural heterogeneities polarized in an external electromagnetic field.[1-2] Full understanding the atomic-scale structural origin of these dipoles is extremely important, however it is limited by experimental inaccessibility. The broader question, namely how individual structural sites interact with an electromagnetic field, could provide new insight into the fundamental physics of solid dielectrics and lead to the development of high-performance dielectric materials for technological applications.

Carbon allotropes and their derivatives can be endowed with multiple dipolar configurations at the atomic scale by incorporating vacancies and hetero-atoms into the graphitic layers.[3-6] Among these findings, transition-metal atoms at doping sites have been proven to produce considerable changes in the local electronic structure and result in significant physical phenomena including spintronics, plasmonics, magnetics and dielectrics. Using ab initio calculations in a hybrid Co/carbon nanotube structure, Yang *et al.* found an enhanced magnetic moment and spin polarization at the Fermi level.[7] Costa *et al.*[8] and Kirwan *et al.*[9] confirmed theoretically various types of exchange magnetic exchange coupling between metal atoms and carbon nanotubes, demonstrating that the interaction range depends on the position and nature of the foreign atoms. Using electron energy-loss spectrum imaging technology, Zhou *et al.* observed plasmon resonances of a single silicon atom substituted in a graphene layer, induced by collective oscillations of the electron density at C/Si interfaces, where the individual silicon atom acts as an antenna in the petahertz ($10^{15}$ Hz) range.[6] Additionally, such atomic-scale substitutions can also result in differences, for example, the enhanced dielectric losses at 8-12 GHz for Boron-doped carbon nanotubes[10] and the cryogenic electromagnetic absorption at 9 GHz for Potassium- and Calcium-doped fullerenes.[11-12] Despite these achievements, overall understanding of the correlation between macroscopic electromagnetic behaviors and the atomic-scale microstructures, particularly for the single-atom configurations, is still relatively poor.

Very recently, we synthesized a (Nb+NbC)@C composite by a modified arc-discharge method, in which single-atom niobium can be stabilized into graphitic layers at high density.[13] Both theoretical and experimental studies showed the most probable atomic configuration and a significant electronic interaction between the single-atom niobium and the surrounding carbon atoms. This development provides a new opportunity to study the intrinsic electromagnetic behaviors of single-atom niobium substituted into graphitic layers. In this work, we experimentally observed a strong dielectric polarization phenomenon at ~16 GHz for the single-atom Nb@C complex, pointing to the existence of permanent dipoles induced by atomic-scale charge interactions at the Nb@C heterogeneity. Together with the nanoscale interfacial polarization arising from the NbC@C interfaces, we observe increased complex permittivity in the (Nb+NbC)@C composite across the whole 2-18 GHz range and enhanced dielectric loss at 11-18 GHz.

A fifth-order aberration-corrected scanning transmission electron microscope (STEM, Nion-UltraSTEM100) operated at an accelerating voltage of 60 kV, coupled with an electron energy-loss spectrometer (EELS, Gatan Enfina spectrometer) was employed to study the microstructure of the (Nb+NbC)@C composite. An accelerating voltage of 60 kV, which is below the knock-on damage threshold of graphene, provides non-destructive structural examination of the single-atom niobium. As observed in **Fig. 1a** and **b**, the carbon matrix was composed mainly of ultra-small NbC clusters with a mean diameter of ~1.6 nm. The high-magnification image in **Fig. 1c** provides clear evidence of high-density single-atom niobium (about $3.4 \times 10^5$ niobium/μm$^2$) uniformly dispersed in the graphitic layers. A ~2 nm NbC cluster encased in graphitic shells is directly identified in a magnified annular dark-field (ADF) image and the corresponding composition visualization is given in **Fig. 1d-f**.

The real part (*ε'*) and imaginary part (*ε''*) of the **(Nb+NbC)@C** composite, shown in **Fig. 2a**, present an overall decline from 23.35 to 11.56 and 16.07 to 7.58, respectively, over the 2-18 GHz frequency range. It should be noted that two obvious resonances appear at ~11 and ~16 GHz. In comparison, the complex permittivity of NbC@C composite (Supplementary **Figure S1**) with a much-reduced concentration of single-atom niobium and the larger niobium carbide clusters only exhibits a resonance at ~11 GHz (**Fig. 2b**). Furthermore, the attenuation coefficient (α) can be calculated from the complex permittivity according to Eq. (1).[17] As shown in **Fig. 2c and d**, both the dielectric loss factor and attenuation coefficient of (Nb+NbC)@C composite present two overlapped resonances, while the NbC@C only exhibits a resonance behavior at ~11 GHz.

$$\alpha = \frac{2\pi f}{c}\left[\frac{\left(\varepsilon'^2+\varepsilon''^2\right)^{1/2}-\varepsilon'}{2}\right]^{1/2} \qquad (1)$$

The dielectric resonance is usually induced by the electron polarization, the ion polarization, and the electric dipolar polarization.[14] At microwave frequencies, the first two polarizations are relatively weak as their dielectric resonances appear at the infrared or higher frequencies. Therefore, the resonances for the (Nb+NbC)@C and NbC@C composites are attributed to the existence of dipolar polarization. Dipolar polarization results from permanent dipoles that arise from the asymmetric charge distribution at heterogeneous interfaces. Under an external electromagnetic field, this polarized charge oscillates at a specific timescale, resulting in a delayed response to the external field (Supplementary **Figure S2** and the discussion).[15-17]

To heterogeneous microstructures, the oscillation of localized space charges at sub-nanometre scale interfaces is, therefore, considered a main contributor to the dielectric polarization phenomenon with similarities to a boundary layer capacitor.[14-16, 18-21] Assuming all the NbC clusters are spherical in shape, one can estimate that the increased interface area for the (Nb+NbC)@C composite is about 30 % higher relative to the NbC@C composite. The complex permittivity of the (Nb+NbC)@C composite is enhanced by 50% relative to the NbC@C composite across the studied frequency range. Interfacial polarization occurs when the motion of these migrating charges is impeded, and trapped within the interfacial region, the so-called Maxwell-Wagner-Sillars (MWS) polarization[22-23]. MWS usually occurs in the megahertz frequency range, causing a significant variation in permittivity. If the charge layers are much smaller than the interfacial dimensions, the charge responds independently of the charge on nearby materials (particles). In this case, the charges have time to accumulate at the borders of the materials, resulting in the interfacial polarizations. However, it should be emphasized, that for the NbC@C composite, the particle sizes range from atomic-scale Nb-C complex to ~2 nm NbC clusters, which might be smaller than the charge displacement.

To reveal the origin of the interfacial polarizations, we performed intensive numerical simulations using the finite element solver COMSOL Multiphysics. The calculated permittivity (real parts) for the simulated NbC@C core@shell model (Supplementary **Figure S3a**) is plotted in **Figure S3b** and shows the classic dielectric resonance at around $10^{12}$ Hz for graphitic shell conductivity set to $3\times10^3$ S/m (the approximate value for the bulk graphite). Clearly the conductivity of graphitic shells can be seriously degraded due to high defect density of substitutional niobium or carbon vacancies, [24-25] and are seen in microstructural characterizations. By choosing the conductivity of graphitic shells to be 3 S/m, the simulated dielectric resonance frequencies can be adjusted to be ~$10^{10}$ Hz, to match our experimental results (~11 GHz). **Figures S3c** shows the electric field distributions of the NbC@C nanoparticle for different graphitic shell conductivities at frequencies of $10^8$, $10^{10}$ and $10^{12}$ Hz, respectively. At these resonance frequencies, the electric fields rapidly change in the core@shell interfaces, providing the direct view of the interfacial polarizations. Combined with the experimental results and theoretical simulations, the observed resonance at ~11 GHz is attributed to the co-operative effect of the core@shell configuration of NbC@C particles and the degraded conductivity of highly defective graphitic shells.

In addition to the interfacial polarization, the (Nb+NbC)@C composite presents another polarization at ~16 GHz. The macroscopic dielectric polarization and loss are ascribed to the existence of permanent dipoles, which could also be achieved by heteroatom doping, for example, K and Ca- doped fullerenes at cryogenic temperature around ~9 GHz,[11-12] Li-/Ti-doped oxides at megahertz frequencies [26-28] and nitrigen-doped carbon nanotubes at near-infrared frequencies.[29-30] To understand the formation of a permanent polarization center for a single-atom niobium, **Fig. 3a** shows an ADF image focused on a magnified region of single niobium atoms in graphitic layers. Careful observation reveals that these niobium single-atoms occupy substitutional sites in carbon atom planes. ADF intensity mapping, associated with atomic numbers, were obtained by using ImageJ digital processing and are shown in **Fig. 3b** and **c**. Localized ADF enhancement relative to the graphitic layer background appears at the single-atom niobium level. The ADF intensity line profile, marked in **Fig. 3a**, was evaluated quantitatively in **Fig. 3d**. The signal-to-background ratio of ADF intensities for the single-atom niobium to graphitic layers is increased by a factor of 6.64 (**Fig. 3e**). Furthermore, **Fig. 3f**, **g,** and **h** present the average *d*-spacing distances of graphitic layers of two representative regions with and without the effect of single-atom niobium. The mean (002) plane distances are 0.351 nm and 0.368 nm, respectively, verifying that the substituted niobium atom results in lattice expansion of the graphitic layers. [13] The atomic-scale characterization suggests that the substitutional single-atom niobium induces a symmetry breaking in the microstructures and provides a possible explanation for the origin of the electric dipolar responsible for dielectric resonance at ~16 GHz.

To further explore the single-atom niobium polarization, we performed a first-principles calculation using a representative single-atom niobium site from the most probable configuration, as shown in **Fig. 4a**. The most energetically favorable configuration is theoretically estimated to be a single-atom niobium substituted into a tri-vacancy site in a carbon atom plane.[13] In this case, the charge flows from the niobium atom (blue) into the C atoms (yellow), as indicated by the top and side-view iso-surfaces of the difference in local electron density distributions shown in **Fig. 4b** and **c**. The consistency between the experimental observations and theoretical predictions strengthens our hypothesis that the single-atom niobium substitution gives rise to a local structural asymmetry and results in the formation of a permanent electric dipole which can excite an oscillation under a specific frequency (**Fig. 4d**).

These experimental and theoretical results provide direct evidence for the electromagnetic characteristics of single-atom niobium substituted in graphitic layers at microwave frequencies. The specific electromagnetic resonance indicates that single-atom niobium plays an antenna-like role in the microwave-matter interaction process at the excitation energy of ~$6.62 \times 10^{-3}$ meV (16 GHz) through the vibration of a permanent dipolar arising from symmetry breaking in the localized microstructure. Given the diversity of single-atom metals, similar phenomenon is likely across various frequencies for doped carbon allotropes and their derivatives. From a fundamental perspective of physics, it would also be significant to explore the electromagnetic characteristics of various substitutional single-atom configurations. Further theoretical and experimental efforts are encouraged to fully uncover these atomic-scale electromagnetic phenomena.

The authors gratefully acknowledge the National Natural Science Foundation of China (51471045), the start-up funding supported from the Northeastern University of China and the Program for Changjiang Scholars, and Innovative Research Team in University (IRT0713). This research used facilities provided by Oak Ridge National Laboratory's ShaRE User Facility sponsored by the Scientific User Facilities Division, Office of Basic Energy Sciences, U.S. Department of Energy. P.F.G. thanks the Center for Computational Materials Science, Institute for Materials Research, Tohoku University, for providing us the Hitachi SR11000 (model K2) supercomputing system. The authors sincerely appreciate the significant discussion from Dr. Jamal Daoud, Dr. Keith Morton, and Dr. Matthew F. Chisholm.

**Figure captions**

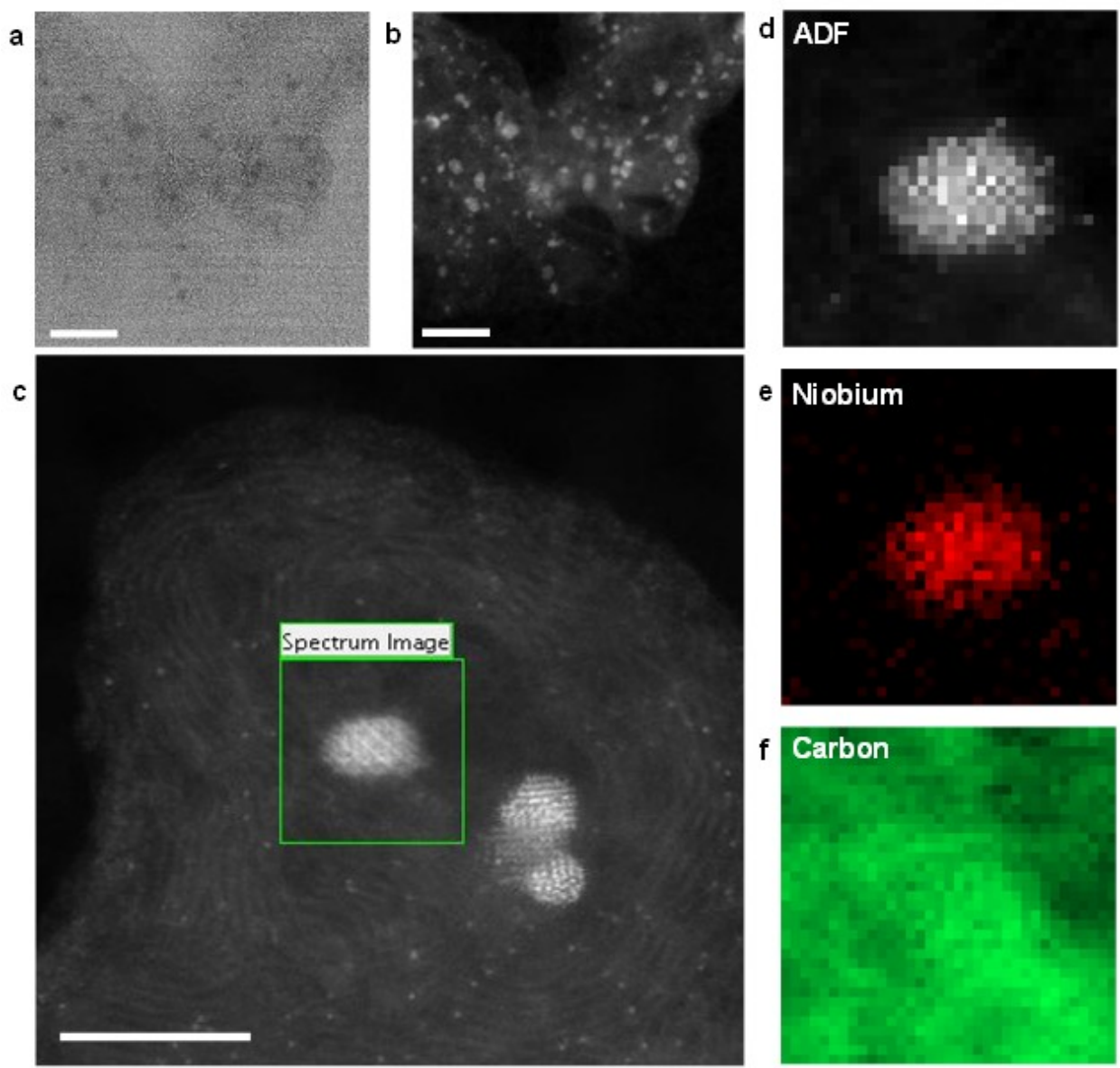

**Figure 1. Microstructures of (NbC+Nb)@C nanocomposite consisting of ultra-small niobium carbide clusters and a high density of single-atom niobium in graphitic layers. a and b**, TEM and STEM images (Scale bar: 10 nm); **c**, High-magnified STEM image, showing ~2 nm niobium carbide clusters and a high density of single-atom niobium doped into onion-like graphite shells (Scale bar: 5 nm); **d-f**, a ADF image of the marked region (green square) in **Fig.1c** and the corresponding EELs mapping images of niobium and carbon elements.

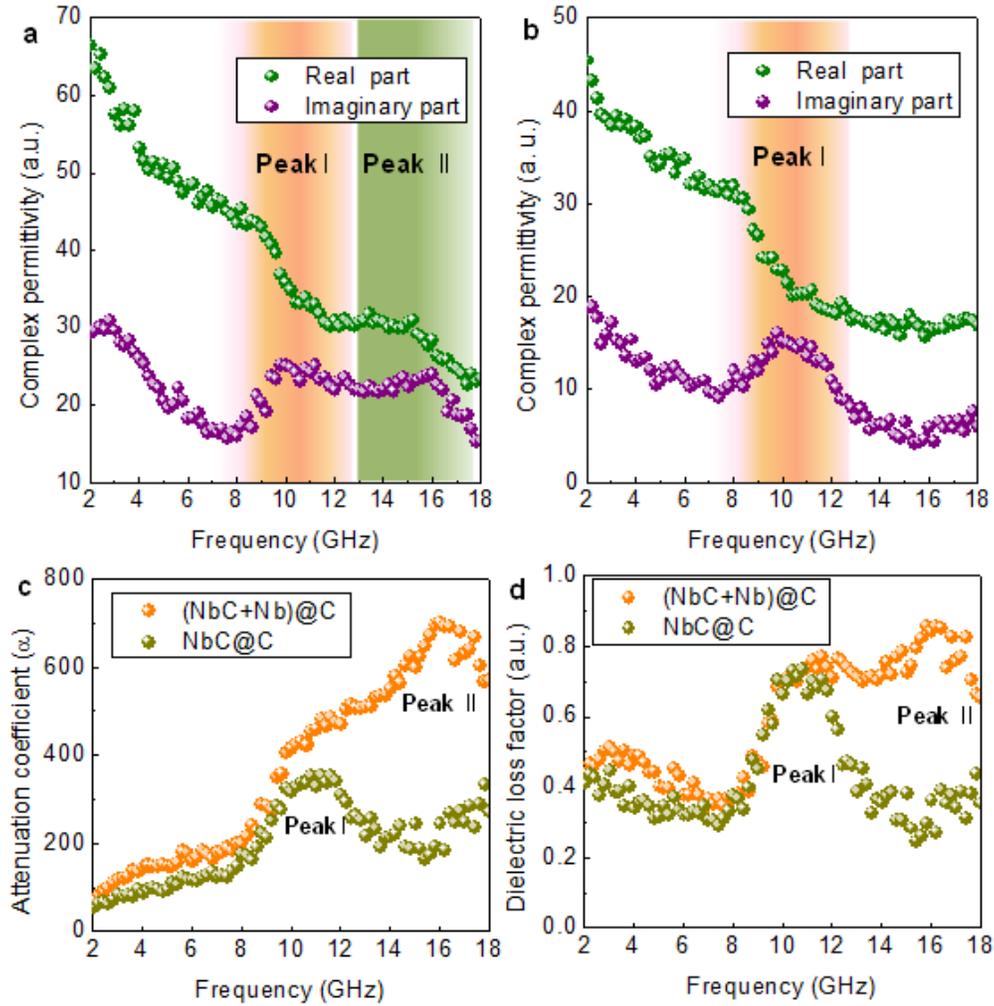

**Figure 2. Electromagnetic properties of (NbC+Nb)@C and NbC@C nanocomposites. a** and **b**, Real part and imaginary part of complex permittivity of (NbC+Nb)@C and NbC@Cnanocomposites as a function of frequency, respectively; **c**, The attenuation coefficient plots of (NbC+Nb)@C and NbC@C nanocomposites. **d**, The dielectric loss factor plots of (NbC+Nb)@C and NbC@C nanocomposites. In the (NbC+Nb)@C nanocomposite, the attenuation coefficient and dielectric loss factors present two peaks (I and II) at around 11 and 16 GHz, while only one peak appears in the NbC@C nanocomposite. Such a control experiment evidently indicates that the origin of the peak II at 16 GHz is ascribed to the single-atom niobium.

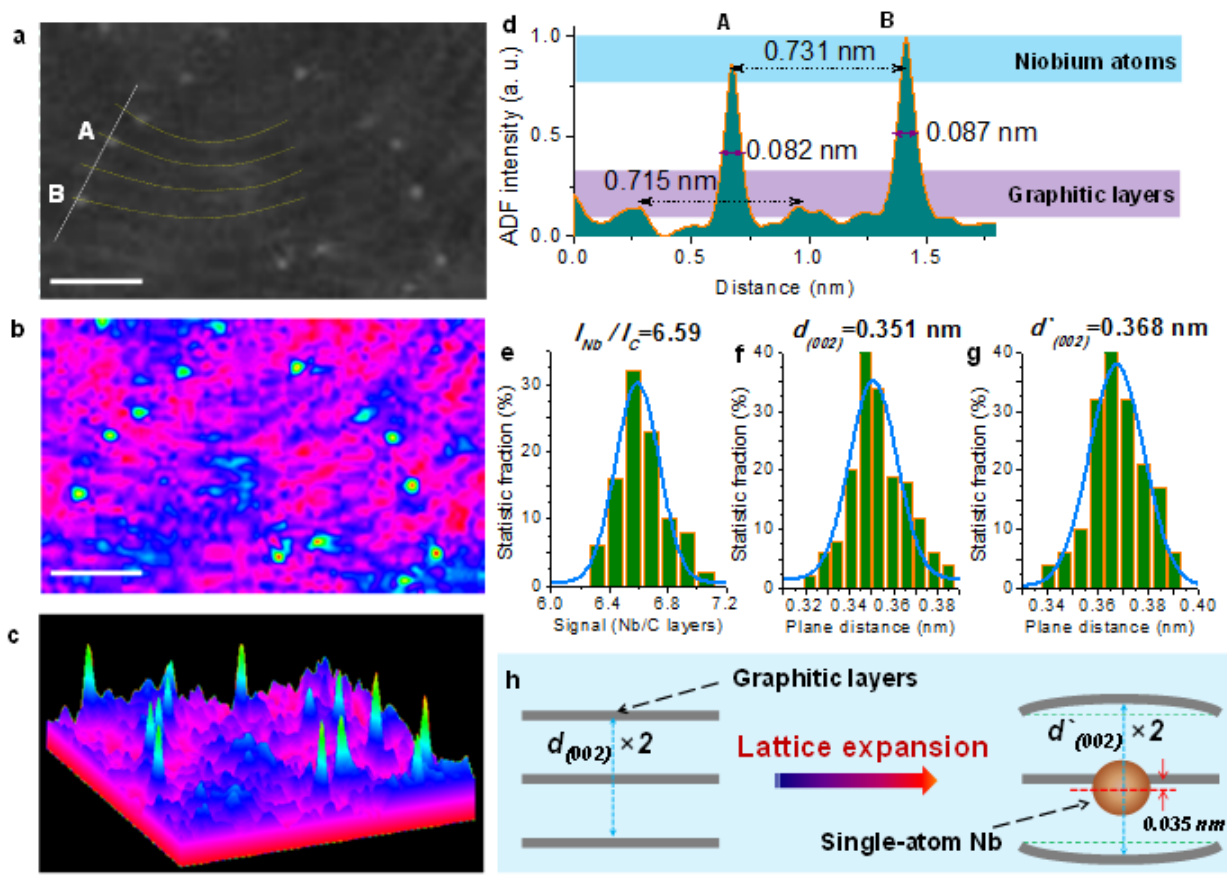

**Figure 3. Configuration of single-atom niobium in graphitic layers. a** and **b**, STEM-ADF Z-contrast images of a high density of single-atom niobium in graphite layers (Scale bar: 1 nm); **c**, High-magnified 3-dimensional view indicated the ADF intensity of the single-atom niobium and the graphitic layers. **d**, ADF intensity profiles of two marked single-atom niobium (A and B) in **a**; **e**, The statistic ADF intensity ratio of the single-atom niobium to the carbon atoms of graphitic layers from more than 200 measurements. **f-g**, The statistic (002) plane distances of graphitic layers at two representative regions shown in the left and right panels in **h**, respectively.

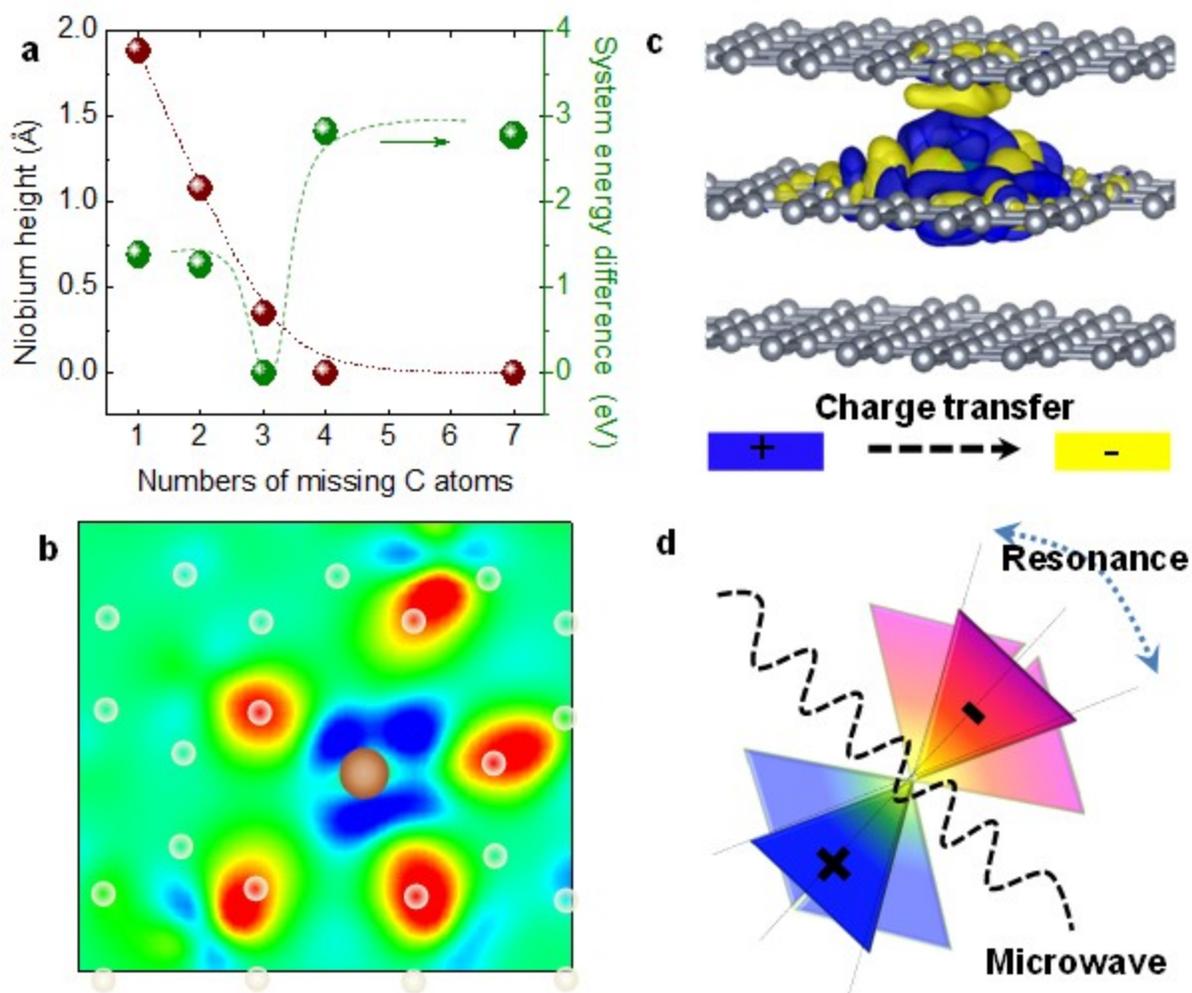

**Figure 4. Origin of the dielectric polarization of single atom niobium in (NbC+Nb)@C nanocomposite. a**, Theoretical simulations of possible structure configurations and system energy differences of a single niobium atom incorporated into various defects on a single-layer carbon plane with *n* missing carbon atoms in a vacancy; **b** and **c**, Top and side- views of the electron density configuration at the vacancy of 3 missing carbon atoms, showing the asymmetric charge distribution at the region of single-atom niobium site, forming a permanent electrical dipolar. Charge flows from the niobium atom (blue) into the C atoms (yellow). **d**, Schematic of the dielectric resonance of an electric dipolar under an external microwave field.

**Supplementary information**

1. **Electromagnetic parameter measurement**

The as-made (Nb+NbC)@C powders were dried in a vacuum and then mechanically mixed with an equal weight of paraffin wax (chosen for its negligible electromagnetic response.). For complex permitivity measurements the mixture was compacted and cut into toroidal-shaped samples of 7.0 mm outer diameter, 3.0 mm inner diameter, and 1.0 mm thickness. Complex permittivity measurements were recorded from 2-18 GHz using an Agilent 8722ES vector network analyzer (VNA) with a sweep oscillator. The VNA was calibrated for full, two-port measurements for reflection and transmission at each port and then connected to the coaxial line sample holder. As a reference, we also performed the same measurements for another composite [denoted as NbC@C hereafter] that was synthesized at a relatively moderate cooling condition consisting of ~2 nm niobium carbide clusters dispersed in graphitic shells. The absence of single-atom niobium in the NbC@C composite enables us to determine unambiguously the origins of electromagnetic characteristics of the (Nb+NbC)@C composite.

2. **Theoretical simulations**

To reveal the origin of the interfacial polarizations, we performed intensive numerical simulations using the finite element solver COMSOL Multiphysics. Supplementary **Figure S3a** shows the geometric model of a core@shell NbC@C nanoparticle with 0.75 nm NbC core diameter and 0.375 nm graphitic shell thickness. The NbC@C nanoparticle is encased in a 4×4×4 nm paraffin matrix, which is insulated on each side. The block is grounded below as a bottom electrode (V=0 voltage) and a potential applied above as a top electrode with V=1 voltage.

**Supplementary figures**

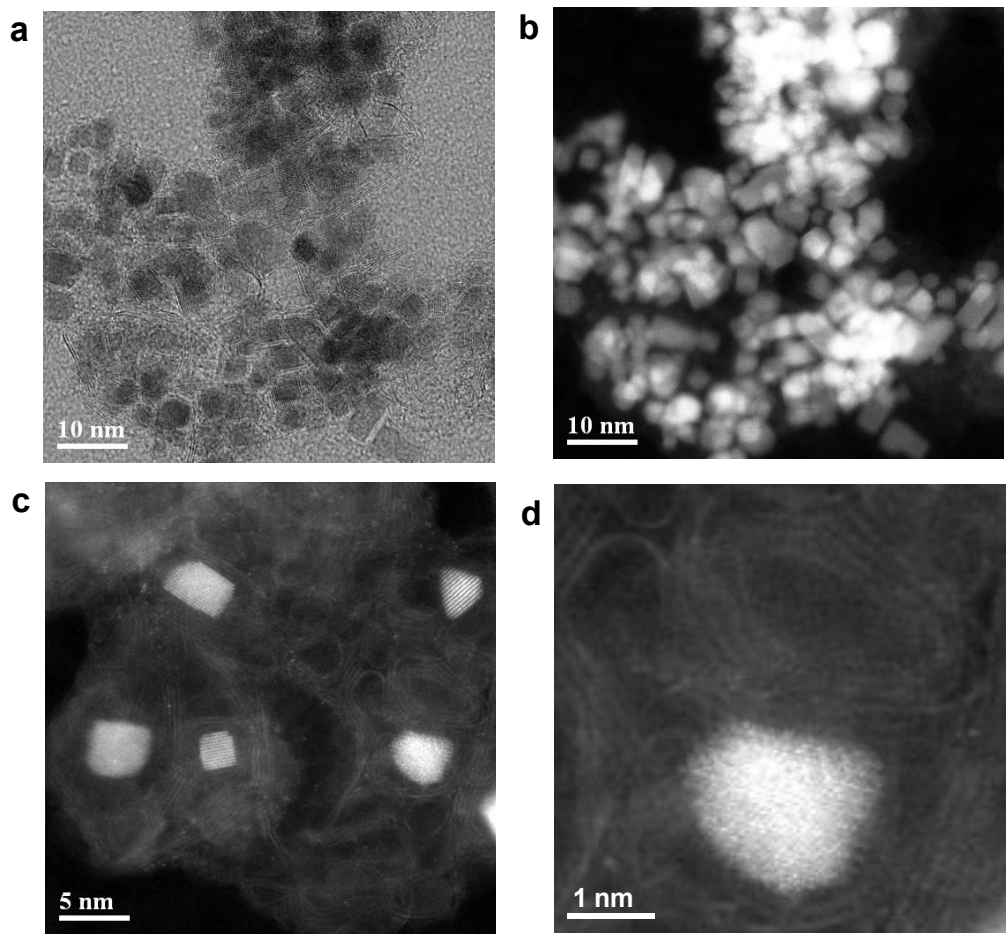

**Figure S1** TEM and ADF images of NbC@C nanocomposite consisting of ultra-small niobium carbide clusters and a much reduced number of incorporated single-atom niobium in graphitic layers.

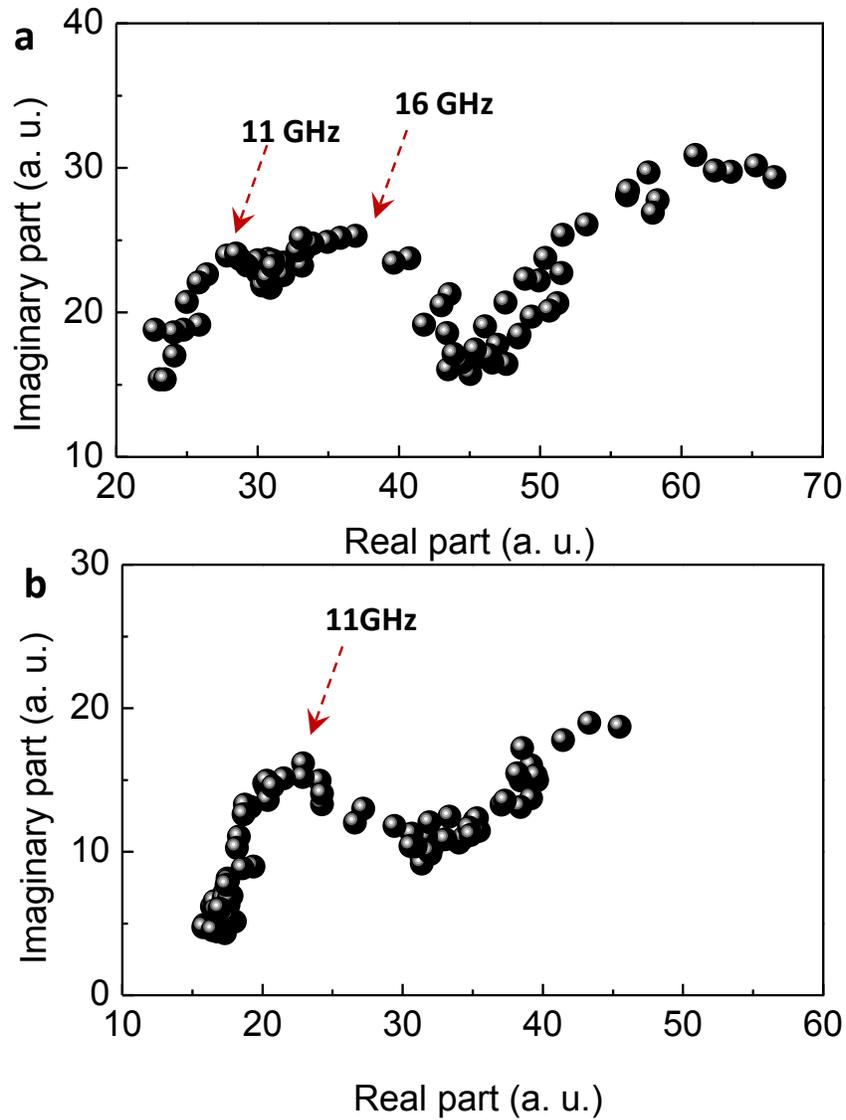

**Figure S2** The Cole-Cole plots of the complex permittivity real part and imaginary part independent of the frequency, in which the (Nb+NbC)@C composite presents an approximate relationship of two overlapped semicircles at ~11 GHz and ~16 GHz while the NbC@C composite exhibits only a semicircle corresponding to the region of ~11 GHz.

The dipolar polarization process can be expressed approximately by the Debye relaxation[17]:

$$\varepsilon_r = \varepsilon_\infty + \frac{\varepsilon_s - \varepsilon_\infty}{1 + j2\pi f \tau} = \varepsilon'(f) + i\varepsilon''(f) \tag{S1}$$

where $f$ is the frequency of the electromagnetic wave, $\tau$ is the relaxation time, $\varepsilon_s$ and $\varepsilon_\infty$ are the stationary and optical dielectric constants, respectively. From Eq. (S1), Eq. (S2) and Eq. (S3) can be deduced:

$$\varepsilon'(f) = \varepsilon_\infty + \frac{\varepsilon_s - \varepsilon_\infty}{1 + (2\pi f)^2 \tau^2} \tag{S2}$$

$$\varepsilon''(f) = \frac{2\pi f \tau (\varepsilon_s - \varepsilon_\infty)}{1 + (2\pi f)^2 \tau^2} \tag{S3}$$

$$\left[\varepsilon' - \frac{\varepsilon_0 + \varepsilon_\infty}{2}\right]^2 + (\varepsilon'')^2 = \left(\frac{\varepsilon_0 + \varepsilon_\infty}{2}\right)^2 \tag{S4}$$

Arranging Eq. (S2) and (S3) yields Eq. (S4), which shows that the plot of $\varepsilon'$ vs. $\varepsilon''$ is a semicircle. The (Nb+NbC)@C composite presents an approximate relationship of two overlapped semicircles at ~11 GHz and ~16 GHz while the NbC@C composite exhibits a single semicircle corresponding to the region of ~11 GHz (Supplementary **Figure S2**).

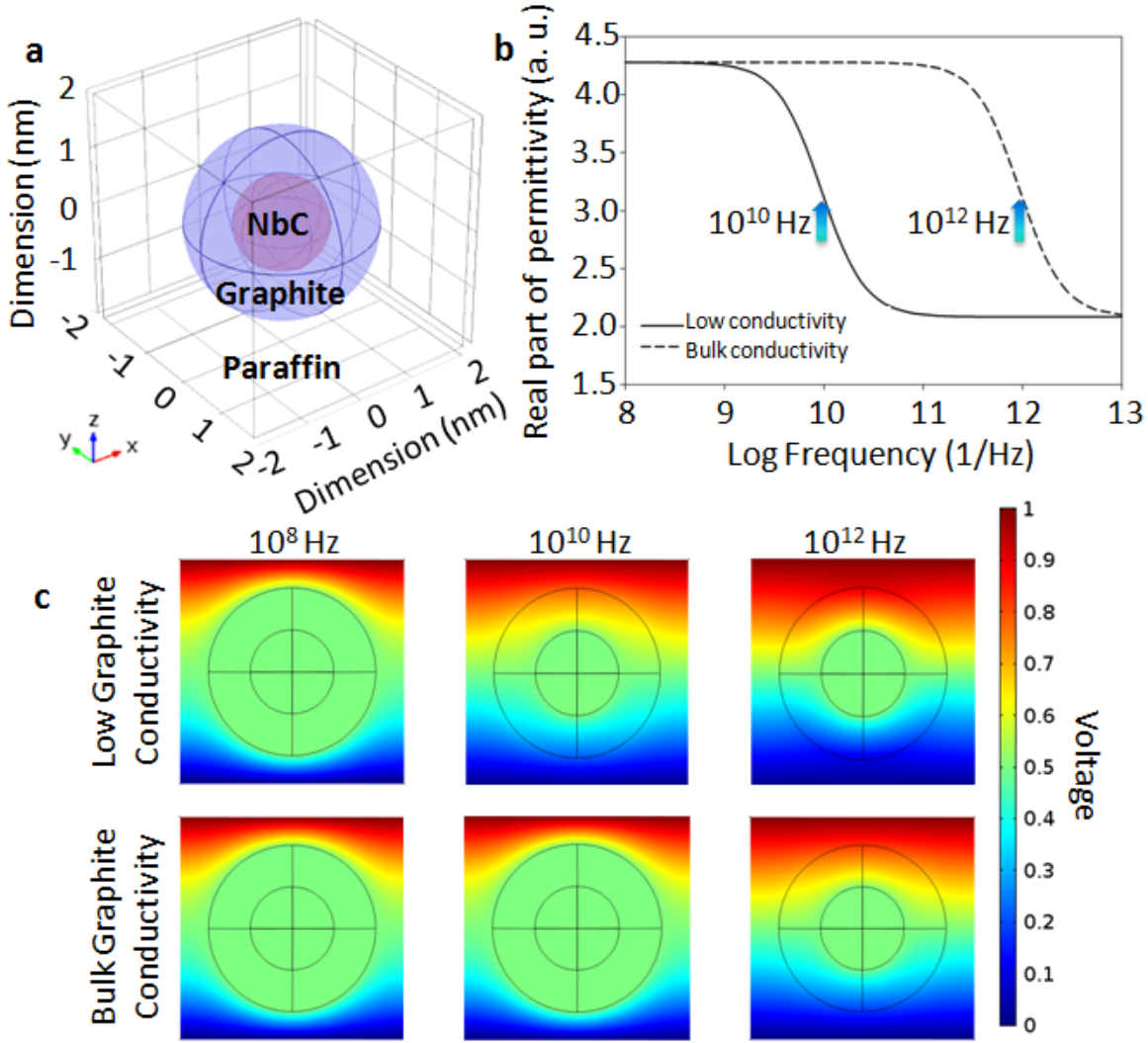

**Figure S3. Dielectric polarization configuration of a NbC@C nanoparticle surrounded by paraffin matrix**. **a**, Geometric model of the core/shell NbC@C nanoparticle consisting of 0.75 nm NbC core in diameter and 0.375 nm graphitic shell in thickness. The NbC@C particle is blocked into a 4×4×4 nm paraffin matrix; **b**, The calculated real parts of permittivity, in which the dielectric resonance frequencies shift down to low frequency as decreasing the conductivity of graphitic shells from $3\times10^3$ S/m to 3 S/m. **c**, the electrical field distributions at frequencies of $10^8$, $10^{10}$, and $10^{12}$ Hz for a NbC@C nanoparticle, evidently demonstrating the core@shell interfacial dielectric polarizations at resonance frequencies.